\documentclass[aps, prl, twocolumn, showpacs,amsmath,amssymb]{revtex4}
\usepackage{graphicx}
\usepackage{bm}
\begin{document}

\title{Modelling the Berezinskii-Kosterlitz-Thouless Transition in the NiGa$_2$S$_4$}
\author{Chyh-Hong Chern}
\email{chern@issp.u-tokyo.ac.jp}

\affiliation{Institute for Solid State Physics, University of
Tokyo, 5-1-5 Kashiwanoha, Kashiwa, Chiba 277-8581, Japan}

\begin{abstract}
In the two-dimensional superfluidity, the proliferation of the
vortices and the anti-vortices results in a new class of phase
transition, Berezinskii-Kosterlitz-Thouless (BKT) transition. This
class of the phase transitions is also anticipated in the
two-dimensional magnetic systems.  However, its existence in the
real magnetic systems still remains mysterious. Here we propose a
phenomenological model to illustrate that the novel spin-freezing
transition recently uncovered in the NMR experiment on the
NiGa$_2$S$_4$ compound is the BKT-type. The novel spin-freezing
state observed in the NiGa$_2$S$_4$ possesses the power-law
decayed spin correlation.
\end{abstract}

 \maketitle

As the thermodynamic conditions in the environment change, for
example the pressure or the temperature, matter transforms from
one state to another.  Steam levitates from the top of the hot
coffee; ice melts in the soft drink.  The phase transition is
ubiquitous in our daily life.  Investigating the phase transition
is always the central subject in physics.  Most of the phase
transitions can be understood by the distinct physical properties
between the phases.  For example, in the vapor-liquid transition,
the unit volumes per mole of the molecules are different from the
vapor to the liquid.  In the ferromagnetic transition, the spins
orientate randomly at the high temperature side but start to align
along the same direction resulting in the net magnetization at the
low temperature side.   However, in the two-dimensional
superfluid, the Berezinskii-Kosterlitz-Thouless (BKT) transition
that happens when vortices and the anti-vortices proliferate does
not separate the phases with distinct thermodynamic
quantities\cite{kosterlitz1973}.  Instead, the superfluidity
correlation changes from the power-law behavior to the
exponentially decayed one as we across the transition from the
lower temperature side. In the two-dimensional magnetic systems,
the BKT transition is also anticipated in all easy-plane
Heisenberg models.  On the other hand, the power-law decayed spin
correlation may play important role in the high transition
temperature superconductor\cite{rantner2001}. Therefore,
understanding the spin dynamics in the critical \emph{phase} has
become tremendously important.

NiGa$_2$S$_4$ is originally synthesized intending to realize the
spin liquid proposed by Anderson few decades
ago\cite{anderson1973}.  It is the layered material that the
spin-1 Ni$^{2+}$ ions form the ideal two-dimensional triangular
network.  As an anti-ferromagnetic insulator, NiGa$_2$S$_4$
exhibits no long-ranged magnetic ordering down to 350mK shown in
the specific heat capacity, the magnetic susceptibility, and the
neutron scattering experiments\cite{Nakatsuji2005}.  At 1.5K, the
Edwards-Anderson order parameter, Q = $1/N\sum_i <S_i>^2$, that
tells the spin moment, is measured at 0.61, where $N$ is the total
number of spins. This vastly reduced spin moment indicates the
presence of the strong quantum fluctuation that is highly
favorable by the scenario of the spin liquid.  However, in the
recent experiments of the Ga nuclear magnetic resonance (Ga-NMR)
\cite{Takeya2007}, a temperature $T_f$ is found around 10K below
which the spin dynamics is slow down and the freezing behavior is
observed.  As approaching the $T_f$ from above, both the nuclear
spin-lattice relaxation rate 1/T$_1$ and the nuclear spin-spin
relaxation rate 1/T$_2$ diverge. Moreover, spins do not freeze
immediately at $T_f$ but persist fluctuating down to 2K found in
the nuclear quadruple resonance measurement (NQR).  Below 2K, the
Ga-NQR spectrum becomes very broad and featureless, which implies
the formation of the \emph{static} inhomogeneous internal magnetic
field. First, intuitively, the static internal magnetic field
occurs when the spins freeze up completely.  In this case, the
Edwards-Anderson order parameter should be close to the quantum
number of the spin angular momentum.   Second, all thermodynamic
quantities change smoothly at the transition temperature $T_f$,
but the 1/T$_1$ and the 1/T$_2$ diverge in the NMR signals.
Therefore, the later NMR experiment apparently looks inconsistent
with the previous measurements.  In this Letter, we shall provide
a consistent picture to compromise all the experimental results.
Most importantly, we illustrate that the novel spin-freezing
transition at T=$T_f$ is the long-sought BKT transition in the
two-dimensional magnetic systems.

\begin{figure}[htb]
\includegraphics[width=0.5\textwidth]{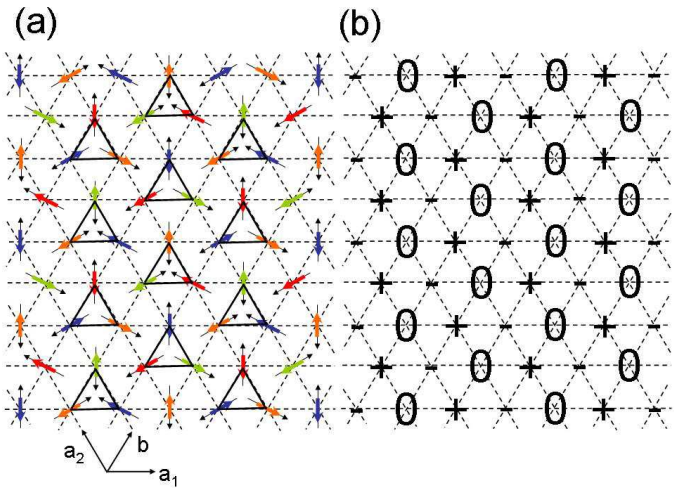}
\caption{(Color online)(a)The spin structure observed by the
neutron scattering in the Ref.\cite{Nakatsuji2005}.  There are two
arrows on every site.  The colored one is the spin orientation,
and the thin black one is the local easy axis. (b) The ground
state of the anti-ferromagnetic quantum Ising model on the
triangular lattice.  "-" indicates the local "-1" state and the
"+" is the local "+1" state.  "0" is the linear superposition of
the "+1" and the "-1" states.}\label{Fig:lattice}
\end{figure}

The way to compromise with all the experiments is to consider the
state that contains both the freezing spins and the fluctuating
ones.  Then, both of them can be observed simultaneously in the experiments.
Before explaining further, let us begin by reviewing the spin
configuration, depicted in Fig.(1a), observed in the neutron
scattering experiment.  The correlation has the wave vector $(1/6,
1/6, 0)$ with the wavelength $2\pi/3a$, where $a$ names the lattice
constant between two Ni$^{2+}$ ions.  This wave vector simply means
that along the {\bf a}$_1$ direction the periodicity is of 6 sites and so
is it along the {\bf a}$_2$, and along the {\bf b} = $(1,1,0)$, the periodicity is
of 3 sites.  For convenience, we highlight the triangles in the
Fig.(1a) as the reference.  With respect to the reference
triangles, spins have the "1-in-2-out" (one spin points in and two
spins point out with respect to the triangles) or the "2-in-1-out" configurations.  Inspired by
the observation in the NMR experiment that the spin correlation starts to develop below
the Curie-Weiss temperature 80K, we assume the existence of an easy
axis on every spin site and its orientation is either parallel or
anti-parallel to the current spin configuration.  In order to
manifest the anti-ferromagnetic nature, we assign the local +z
axis to be the "all-in" or the "all-out" with respect to the
reference triangles alternating over the whole lattice.  An
example of the orientation of the local axes is depicted as the
black thin arrows in the Fig.(1a).  In this transformed
coordinate, spins are either +1 (parallel to the +z direction) or
-1 (anti-parallel to the +z direction).

Considering the quantum fluctuation explicitly, we propose the
following phenomenological Hamiltonian in the transformed
coordinate
\begin{equation}
H=J\sum_{<ij>}\sigma^z_i\sigma^z_j-K\sum_i\sigma^x_i, \label{H}
\end{equation}
where $\sigma^z$ take the eigenvalues +1 and -1, the $J$ is the
positive phenomenological coupling constant, and the $K$ is the
positive measure.  Since the $\sigma^x$ have the off-diagonal
matrix elements that connect the +1 and the -1 states, the $K$
measures the spin-flipping process between the +1 and the -1
states.  For small $K$, the ground state of the Eq.(1) on the
triangular lattice is given in the Fig.(1b).  Using our coordinate
transformation, we can map the spin configuration in Fig.(1a) to
the ground state of Eq.(\ref{H}).  If the spin is parallel to the
local +z-axis, we assign +1; if it is anti-parallel, we assign -1.
In this way, there are spins that are surrounded by the equal
number of the +1 and the -1.  If those spins flip, the energy from
the first term in Eq.(\ref{H}) does not change, but this process
is encouraged by the second term.  Earlier numerical calculation
confirms that eventually the "0 state" is favored, which occupied
+1 and the -1 with the equal probability\cite{Meossner2000,
Meossner2001}.  Then, this state with the thermal fluctuation
naturally resolves the inconsistency between the experiments.  The
Edwards-Anderson order parameter of this state is
0.6677\cite{Blankschtein1984}.  It contains both the freezing and
fluctuating sites.  Below the freezing temperature $T_f$, the
freezing sites contribute to the static internal magnetic field
observed in the Ga-NMR experiment, and the fluctuating sites
remain fluctuating down to zero temperature to contribute to the
reduced moment found in the neutron scattering. We note that
although the spin variable on the "+1" and the "-1" sites may
change with the coordinate transformation, the one on the "0" site
does not.  The Edward-Anderson order parameter is invariant with
respect to our coordinate transformation.

At the mean-field level, as discussed in ref.\cite{Nakatsuji2005},
the state in Fig.(\ref{Fig:lattice}) can be stabilized by a
ferromagnetic nearest neighbor exchange and an anti-ferromagnetic
third-nearest neighbor exchange.  Photoemission spectroscopy also
supports the enhanced contribution from the third-nearest-neighbor
sites\cite{Takubo2007}.  Moreover, in the high temperature, the
superexchange in this material is believed to be isotropic. These
complexity concerning the microscopic details is not included in
this paper.  We remark that our phenomenological model is built on
top of the existence of the abnormal spin correlation below the
Curie-Weiss temperature.  The superexchange in our model is
\emph{anti-ferromagnetic} in the \emph{transformed} coordinate.
Most importantly, the existence of the critical phase, as will be
shown later, and the validity of the current model to describe the
critical behavior help avoid those complexity.  In other words,
Eq.(\ref{H}) might be the fix-point Hamiltonian of the fundamental
microscopic model.

\begin{figure}[htbp]
\includegraphics[width=0.48\textwidth]{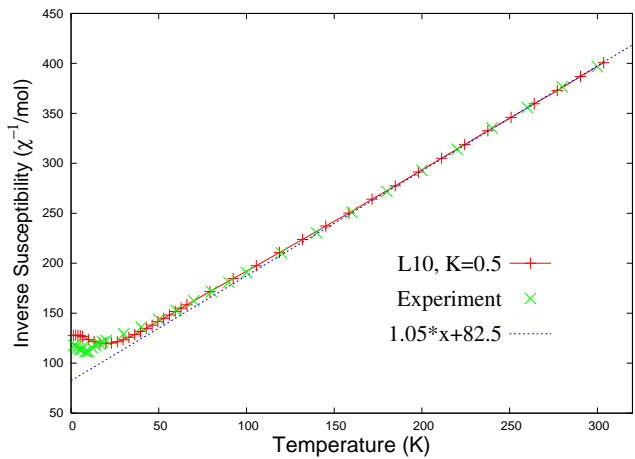}
\caption{(Color online)The system size is $10\times10$ with 100
spins in the calculation.  The theoretical result has the same
structure as the experiments. At the dip of the $\chi^{-1}$,
marked $T_f$, the spin-freezing transition occurs found in the
Ga-NMR experiments.} \label{Fig:sus}
\end{figure}

Now, let us focus on the spin-freezing transition observed at
$T_f$. Since the poly-crystalline sample was used in the magnetic
susceptibility measurement, we apply the quantum Monte Carlo
technique to calculate the averaged susceptibility $\chi$ defined
by $1/3(\chi_{xx}+\chi_{yy}+\chi_{zz})$, where
$\chi_{ij}=dm_i/dh_j$, and $m_i$ is the magnetization per site and
$h_j$ is the external magnetic field.  In the calculation, the
$10^6$ Monte Carlo Steps with the average over 64 ensembles is
used.  In addition, the cluster algorithm is applied along the
imaginary time direction.  In Fig.(\ref{Fig:sus}), the result of
the inverse susceptibility $\chi^{-1}$ with $J=66$K (in the
temperature unit) and $K=0.5J$ is presented.   It is compared
qualitatively well with the experiment and shows weak
size-dependence because of its average nature.  Both of them have
the spoon-like shape that contains the dip, but the position of
the dips happen at different temperature from the theoretical
result and the experimental one.  We mark the temperature of the
dip $T_c$ in the theoretical result and $T_f$ in the experimental
one.

\begin{figure}[htbp]
\includegraphics[width=0.48\textwidth]{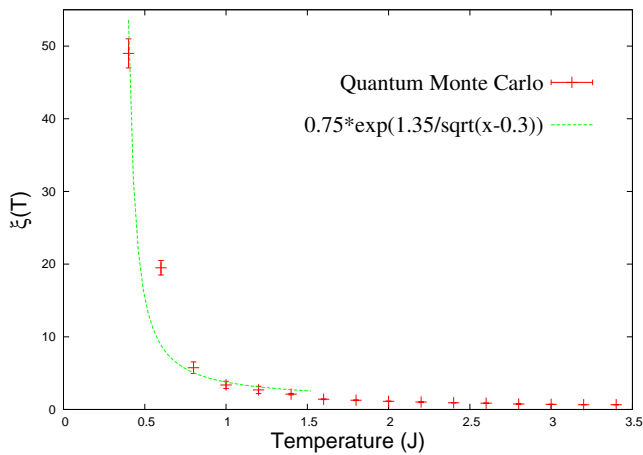}
\caption{(Color online)The result of $\xi(T)$.  The temperature is
in the unit of $J$, and the correlation length is in the unit of
lattice constant $a$. At the high temperature side of $T_c$ $(\sim
0.3J)$, the spin correlation is exponentially decayed, and the
correlation length diverges at $T_c$.  Below $T_c$, the phase has
the power-law spin correlation. The green line is the theoretical
fit of the 2D XY universality class.} \label{Fig:cf_length}
\end{figure}

In the Ga-NMR experiments, the spin-freezing transition is
observed at the dip of the inverse susceptibility, and an unusual
spin correlation starts to develop at the Curie-Weiss temperature
80K. To understand this, we compute the spin-spin correlation
along the {\bf a}$_1$ direction defined by
\begin{equation}
D(n{\bf a}_1)=\frac{1}{N}\sum_i<\sigma^z_i\sigma^z_{i+n}> \label{correlation}
\end{equation}
in the $L_x\times M$ geometry with the periodic boundary
condition, where $L_x$ is the length along the {\bf a}$_1$
direction and $M$ is the one along the $(1,1,0)$ direction, and
$L_x \sim 4M$ is taken.  In this case, the correlation length
$\xi(T, M)$ scales with $M$.  We define
\begin{equation}
\xi(T)= \lim_{M\rightarrow \infty} \xi(T,M),
\end{equation}
and it is shown in Fig.(\ref{Fig:cf_length}).  We find that the
spin correlation starts to develop at $1.4J$ and it diverges at
$T_c\sim 0.3J$.

 \begin{figure}[htbp]
\begin{minipage}{1.0\linewidth}
\includegraphics[width=1.0\textwidth]{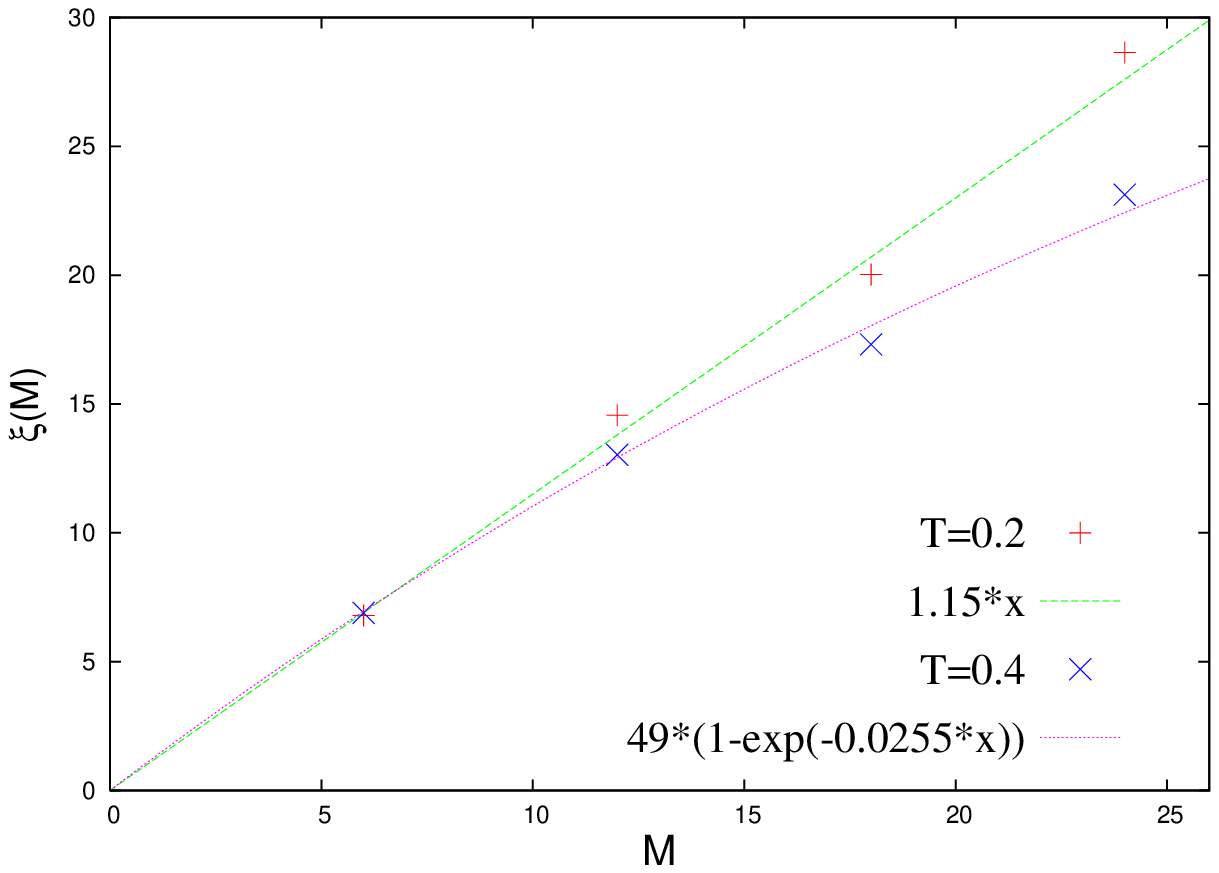}
\caption{(Color online)$\xi(T,M)$ at T=0.2$J$ and $0.4J$. The
vertical axis is the $\xi(T,M)$, and $M$ is in the unit of the
lattice constant $a$. At T=0.4$J$, the correlation length
saturates exponentially. At T=0.2$J$, the linear scaling implies
the critical phase as explained in the text. Two lines are the
functional fit to the data} \label{Fig:24com}
\end{minipage}
\begin{minipage}{1.0\linewidth}
\includegraphics[width=1.0\textwidth]{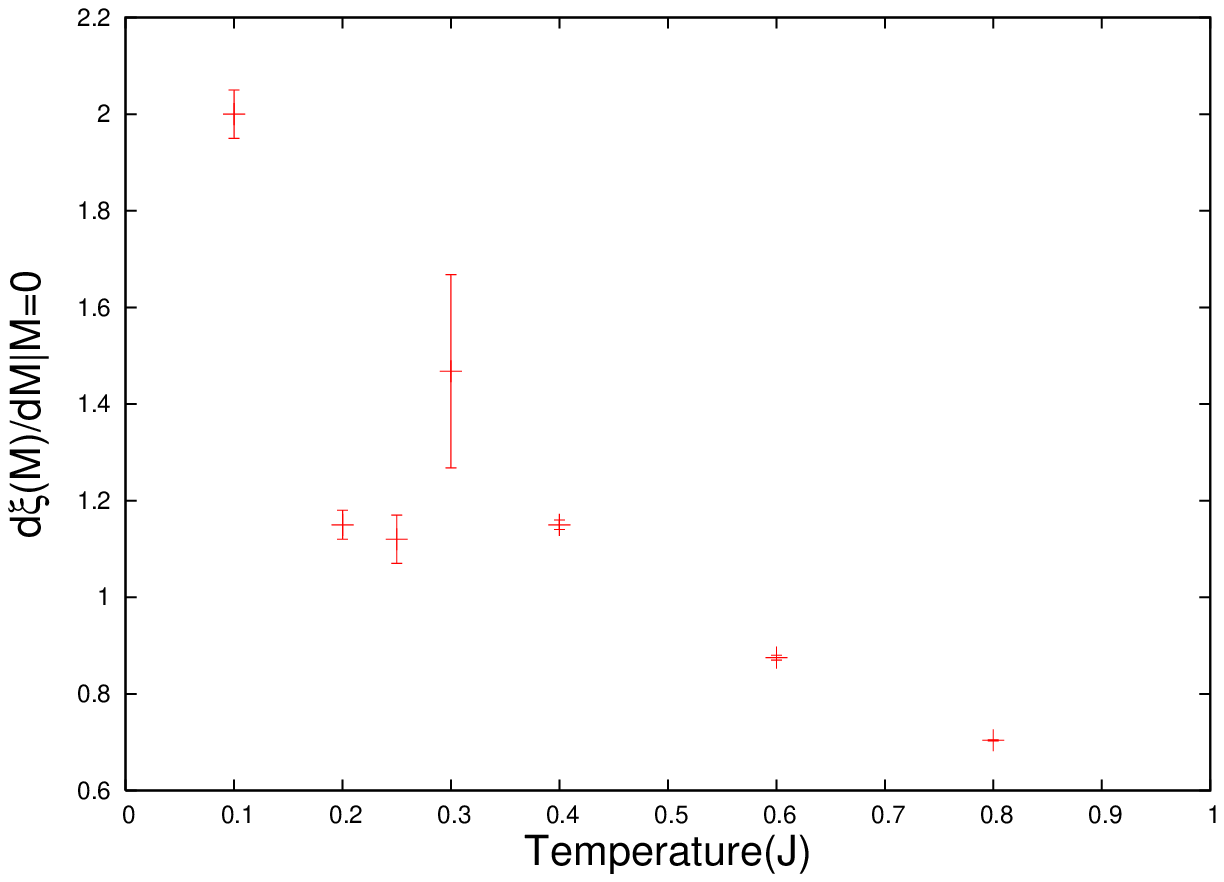}
\caption{(Color online)The temperature is in the unit of $J$.  The
vertical axis is the initial slope of $\xi(T,M)$ defined in the
text. If there is no phase transition, the initial slope should be
exponentially and monotonically increasing. However, there is a
discontinuity at T = 0.3$J$, which indicates a phase
transition.}\label{Fig:cf_slope}
\end{minipage}
\end{figure}

At $T_c$, the BKT-type transition occurs.  Above $T_c$, for
example T=0.4$J$, $\xi(T, M)$ saturates exponentially shown in the
Fig.(\ref{Fig:24com}).  Below $T_c$, for example T=0.2$J$, $\xi(T,
M)$ is linear to $M$.  This linearity has only two possibilities:
1) the correlation length is so large that our system sizes are
too small to reach the saturation.  2) it is in the critical
phase.  Because of the conformal invariance of the critical phase
in the two dimensions, the spin correlation decays exponentially
in the torus geometry.  In this case, $\xi(T, M)$ is linear to M.
Furthermore, the scaling dimension $\Delta(T)$ can be obtained by
\begin{equation}
\xi(T)=\frac{1}{2\pi\Delta(T)}M \label{critical_dim}
\end{equation}
where $\Delta$ is defined by
\begin{equation}
<\sigma^z(0)\sigma^z(r)>\sim\frac{1}{r^{2\Delta}}
\end{equation}
The first possibility can be ruled out as the following.  In
Fig.(\ref{Fig:cf_slope}), we plot the initial slope of $\xi(T,M)$
defined by $\partial\xi(T, M)/\partial M|_{M=0}$.  It shows a
clear phase transition at T=$T_c$.  If the first possibility were
true, the initial slope would have been monotonic and grew
exponentially as temperature decreases.  However, the slope has
the discontinuity at $T_c$, indicating a phase transition.
Moreover, it is not a second-order phase transition, because both
the magnetic susceptibility and the specific heat are smooth
functions at $T_c$.  If it is a classical Ising transition,
$\xi(T)$ should be symmetric with respect to $T_c$ as
$|T-T_c|<<1$.  Here, although the initial slopes at T=0.2$J$ and
T=0.4$J$ are the same within the error bar, their asymptotic
behaviors are entirely different as shown in
Fig.(\ref{Fig:24com}). Therefore, the phase below $T_c$ should be
critical with the power-law spin correlation. Additionally, the
scaling dimension $\Delta(T)$ has the monotonic temperature
dependence, which is the typical behavior of the scaling dimension
in the critical phase. Due to the restriction of the technique, we
are not able to compute the free energy to find the central charge
in the critical phase. Classification by the conformal field
theory may be an interesting directions for the future research.

The existence of the BKT transition in the quantum Ising model on
the triangular lattice was previously pointed out by R. Moessner,
et al.\cite{Meossner2001}.  Here we summarize their argument and
further construct the topological object in this model.  The
quantum 2D model of Eq.(\ref{H}) at finite temperature can be
mapped to a 3D classical Ising model with the ferromagnetic
exchange along the imaginary time direction with \emph{finite
dimension}. Using the Landau-Ginzburg-Wilson analysis, the 3D
model can be mapped to an XY model with a sixth-order symmetric
breaking term which has the sixfold clock symmetry.  At zero
temperature, there is a quantum phase transition described by the
3D XY universality class at finite $K$, because the clock term is
dangerously irrelevant in 3D.  However, at \emph{finite}
temperature and in the thermodynamic limit, the 3D model crosses
over to the 2D model and it results in two transitions: one is BKT
transition at higher temperature and the other at lower
temperature corresponds to the sixfold clock symmetry breaking,
for example, to the state in Fig.(\ref{Fig:lattice}b). Therefore,
the BKT transition in NiGa$_2$S$_4$ actually belongs to the 2D XY
universality class! In Fig.(\ref{Fig:cf_length}), we fit the
correlation result well with the 2D XY model\cite{Jose1977}. These
two transitions are both seen in the Ga-NMR
experiment\cite{Takeya2007}.  The spin-freezing transition at 10K
is the BKT transition, and the transition at 2K where spins
completely freeze up corresponds to the second transition.


It is not coincident that the phase transition occurs at the dip
of the $\chi^{-1}$.  Once the power-law spin correlation survives
in the \emph{phase} rather than a critical point, the response to
the external field could be weaker.  Due to the slow spin
dynamics, the magnetic susceptibility begins to drop in the
critical phase.

\begin{figure}[htbp]
\includegraphics[width=0.48\textwidth]{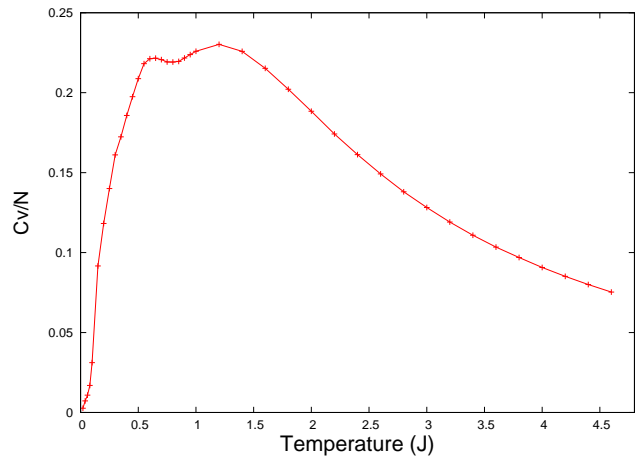}
\caption{(Color online) The temperature dependence of the specific
heat capacity.  The temperature is in Kelvin.} \label{Fig:cv}
\end{figure}

In Fig.(\ref{Fig:cv}), we show the calculation of the specific
heat capacity.  It illustrates the double-peak structure and the
reason is similar to the one given in the Ref.\cite{chern2007}.
Although the valley between the peaks is not as deep as the
experimental one shown in Ref.\cite{Nakatsuji2005}.  The magnetic
specific heat in Ref.\cite{Nakatsuji2005} is obtained by
subtracting the specific heat of ZnIn$_2$S$_4$, which is
non-magnetic and iso-structural to NiGa$_2$S$_4$, from the one of
the NiGa$_2$S$_4$.  We remark that there is 29.68\% difference in
the total atomic mass between these two compounds.  How reliable
their magnetic specific result is suspicious to us.  However,
their low-temperature result may be correct.  Here we also find no
evidence of the existence of the energy gap, which is consistent
with their experiment.

In summary, we have shown that the novel spin-freezing transition
seen in the Ga-NMR experiment on the NiGa$_2$S$_4$ compound is the
BKT-type transition which belongs to the 2D XY universality class.
The divergence of the spin correlation leads to the divergence of
the 1/T$_1$ and 1/T$_2$. Below $T_f$, the power-law spin
correlation develops in the state in the Fig.(1). The truly
long-ranged spin correlation happens at the zero temperature,
dubbed by the "order by disorder"\cite{Meossner2001,
Meossner2000}.  Through our analysis, NiGa$_2$S$_4$ should be
removed from the candidate list for the spin-liquid ground state.
Finally, the long-sought BKT transition in the quantum spin system
is unexpectedly found.  The new phase accompanying with the novel
phase transition will refresh our understanding of the spin
dynamics in the critical phase.

CHC particularly acknowledges S. Nakatsuji and Y. Nambu for
providing the experimental data and the stimulated discussion.  He
deeply appreciates the discussion with M. Oshikawa, with whom the
argument for the BKT transition is formulated.  He is also
grateful for the fruitful discussions with J. Moore and D.
Agterberg.  Most of the calculation is done in the Supercomputer
Center in the Institute for Solid State Physics.


\begin{thebibliography}{11}
\expandafter\ifx\csname
natexlab\endcsname\relax\def\natexlab#1{#1}\fi
\expandafter\ifx\csname bibnamefont\endcsname\relax
  \def\bibnamefont#1{#1}\fi
\expandafter\ifx\csname bibfnamefont\endcsname\relax
  \def\bibfnamefont#1{#1}\fi
\expandafter\ifx\csname citenamefont\endcsname\relax
  \def\citenamefont#1{#1}\fi
\expandafter\ifx\csname url\endcsname\relax
  \def\url#1{\texttt{#1}}\fi
\expandafter\ifx\csname
urlprefix\endcsname\relax\def\urlprefix{URL }\fi
\providecommand{\bibinfo}[2]{#2}
\providecommand{\eprint}[2][]{\url{#2}}

\bibitem[{\citenamefont{Kosterlitz and Thouless}(1973)}]{kosterlitz1973}
\bibinfo{author}{\bibfnamefont{J.}~\bibnamefont{Kosterlitz}} \bibnamefont{and}
  \bibinfo{author}{\bibfnamefont{D.}~\bibnamefont{Thouless}},
  \bibinfo{journal}{J. Phys. C} \textbf{\bibinfo{volume}{6}},
  \bibinfo{pages}{1181} (\bibinfo{year}{1973}).

\bibitem[{\citenamefont{Rantner and Wen}(2001)}]{rantner2001}
\bibinfo{author}{\bibfnamefont{W.}~\bibnamefont{Rantner}} \bibnamefont{and}
  \bibinfo{author}{\bibfnamefont{X.-G.} \bibnamefont{Wen}},
  \bibinfo{journal}{Phys. Rev. Lett.} \textbf{\bibinfo{volume}{86}},
  \bibinfo{pages}{3871} (\bibinfo{year}{2001}).

\bibitem[{\citenamefont{Anderson}(1973)}]{anderson1973}
\bibinfo{author}{\bibfnamefont{P.}~\bibnamefont{Anderson}},
  \bibinfo{journal}{Mater. Res. Bull.} \textbf{\bibinfo{volume}{8}},
  \bibinfo{pages}{153} (\bibinfo{year}{1973}).

\bibitem[{\citenamefont{Nakatsuji et~al.}(2005)\citenamefont{Nakatsuji, Nambu,
  Tonomura, Sakai, Jonas, Broholm, Tsunetsugu, Qiu, and Maeno}}]{Nakatsuji2005}
\bibinfo{author}{\bibfnamefont{S.}~\bibnamefont{Nakatsuji}},
  \bibinfo{author}{\bibfnamefont{Y.}~\bibnamefont{Nambu}},
  \bibinfo{author}{\bibfnamefont{H.}~\bibnamefont{Tonomura}},
  \bibinfo{author}{\bibfnamefont{O.}~\bibnamefont{Sakai}},
  \bibinfo{author}{\bibfnamefont{S.}~\bibnamefont{Jonas}},
  \bibinfo{author}{\bibfnamefont{C.}~\bibnamefont{Broholm}},
  \bibinfo{author}{\bibfnamefont{H.}~\bibnamefont{Tsunetsugu}},
  \bibinfo{author}{\bibfnamefont{Y.}~\bibnamefont{Qiu}}, \bibnamefont{and}
  \bibinfo{author}{\bibfnamefont{Y.}~\bibnamefont{Maeno}},
  \bibinfo{journal}{Science} \textbf{\bibinfo{volume}{309}},
  \bibinfo{pages}{1697} (\bibinfo{year}{2005}).

\bibitem[{\citenamefont{Takeya et~al.}(2008)\citenamefont{Takeya, Ishida,
  Kitagawa, Ihara, Onuma, Maeno, Nambu, Nakatsuji, MacLaughlin, Koda
  et~al.}}]{Takeya2007}
\bibinfo{author}{\bibfnamefont{H.}~\bibnamefont{Takeya}},
  \bibinfo{author}{\bibfnamefont{K.}~\bibnamefont{Ishida}},
  \bibinfo{author}{\bibfnamefont{K.}~\bibnamefont{Kitagawa}},
  \bibinfo{author}{\bibfnamefont{Y.}~\bibnamefont{Ihara}},
  \bibinfo{author}{\bibfnamefont{K.}~\bibnamefont{Onuma}},
  \bibinfo{author}{\bibfnamefont{Y.}~\bibnamefont{Maeno}},
  \bibinfo{author}{\bibfnamefont{Y.}~\bibnamefont{Nambu}},
  \bibinfo{author}{\bibfnamefont{S.}~\bibnamefont{Nakatsuji}},
  \bibinfo{author}{\bibfnamefont{D.~E.} \bibnamefont{MacLaughlin}},
  \bibinfo{author}{\bibfnamefont{A.}~\bibnamefont{Koda}}, \bibnamefont{et~al.},
  \bibinfo{journal}{arXiv:0801.0190, to appear in Phys. Rev. B.}
  (\bibinfo{year}{2008}).

\bibitem[{\citenamefont{Moessner et~al.}(2000)\citenamefont{Moessner, Sondhi,
  and Chandra}}]{Meossner2000}
\bibinfo{author}{\bibfnamefont{R.}~\bibnamefont{Moessner}},
  \bibinfo{author}{\bibfnamefont{S.}~\bibnamefont{Sondhi}}, \bibnamefont{and}
  \bibinfo{author}{\bibfnamefont{P.}~\bibnamefont{Chandra}},
  \bibinfo{journal}{Phys. Rev. Lett.} \textbf{\bibinfo{volume}{84}},
  \bibinfo{pages}{4457} (\bibinfo{year}{2000}).

\bibitem[{\citenamefont{Moessner and Sondhi}(2001)}]{Meossner2001}
\bibinfo{author}{\bibfnamefont{R.}~\bibnamefont{Moessner}} \bibnamefont{and}
  \bibinfo{author}{\bibfnamefont{S.}~\bibnamefont{Sondhi}},
  \bibinfo{journal}{Phys. Rev. B} \textbf{\bibinfo{volume}{63}},
  \bibinfo{pages}{224401} (\bibinfo{year}{2001}).

\bibitem[{\citenamefont{Blankschtein et~al.}(1984)\citenamefont{Blankschtein,
  Ma, Berker, Grest, and Soukoulis}}]{Blankschtein1984}
\bibinfo{author}{\bibfnamefont{D.}~\bibnamefont{Blankschtein}},
  \bibinfo{author}{\bibfnamefont{M.}~\bibnamefont{Ma}},
  \bibinfo{author}{\bibfnamefont{A.~N.} \bibnamefont{Berker}},
  \bibinfo{author}{\bibfnamefont{G.~S.} \bibnamefont{Grest}}, \bibnamefont{and}
  \bibinfo{author}{\bibfnamefont{C.~M.} \bibnamefont{Soukoulis}},
  \bibinfo{journal}{Phys. Rev. B} \textbf{\bibinfo{volume}{29}},
  \bibinfo{pages}{5250} (\bibinfo{year}{1984}).

\bibitem[{\citenamefont{Takubo et~al.}(2007)\citenamefont{Takubo, Mizokawa,
  Son, Nambu, Nakatsuji, and Maeno}}]{Takubo2007}
\bibinfo{author}{\bibfnamefont{K.}~\bibnamefont{Takubo}},
  \bibinfo{author}{\bibfnamefont{T.}~\bibnamefont{Mizokawa}},
  \bibinfo{author}{\bibfnamefont{J.-Y.} \bibnamefont{Son}},
  \bibinfo{author}{\bibfnamefont{Y.}~\bibnamefont{Nambu}},
  \bibinfo{author}{\bibfnamefont{S.}~\bibnamefont{Nakatsuji}},
  \bibnamefont{and} \bibinfo{author}{\bibfnamefont{Y.}~\bibnamefont{Maeno}},
  \bibinfo{journal}{Phys. Rev. Lett.} \textbf{\bibinfo{volume}{99}},
  \bibinfo{pages}{037203} (\bibinfo{year}{2007}).

\bibitem[{\citenamefont{Jose et~al.}(1977)\citenamefont{Jose, Kadanoff,
  Kirkpatrick, and Nelson}}]{Jose1977}
\bibinfo{author}{\bibfnamefont{J.~V.} \bibnamefont{Jose}},
  \bibinfo{author}{\bibfnamefont{L.~P.} \bibnamefont{Kadanoff}},
  \bibinfo{author}{\bibfnamefont{S.}~\bibnamefont{Kirkpatrick}},
  \bibnamefont{and} \bibinfo{author}{\bibfnamefont{D.~R.}
  \bibnamefont{Nelson}}, \bibinfo{journal}{Phys. Rev. B}
  \textbf{\bibinfo{volume}{16}}, \bibinfo{pages}{1217} (\bibinfo{year}{1977}).

\bibitem[{\citenamefont{Chern and Tsukamoto}(2008)}]{chern2007}
\bibinfo{author}{\bibfnamefont{C.-H.} \bibnamefont{Chern}} \bibnamefont{and}
  \bibinfo{author}{\bibfnamefont{M.}~\bibnamefont{Tsukamoto}},
  \bibinfo{journal}{Phys. Rev. B} \textbf{\bibinfo{volume}{77}},
  \bibinfo{pages}{172404} (\bibinfo{year}{2008}).

\end{thebibliography}

\end{document}